\documentstyle[12pt]{article}
\begin{document}

\renewcommand{\thefootnote}{\fnsymbol{footnote}}

\begin{center}

{\large\bf On a Possibility of Membrane Cosmology}\\

\bigskip

 S.N. Roshchupkin \\
 \medskip
 {\it Simferopol State University}\\
{\it 333036, Simferopol, Ukraine}\\

\vspace{5mm}
  and A.A. Zheltukhin\footnote{E-mail:kfti@kfti.kharkov.ua}\\

  \vspace{5mm}

{\it Kharkov Institute of Physics and Technology}\\
{\it 310108, Kharkov, Ukraine}\\
\end{center}
\vspace{1.5cm}

\begin{quotation}
{\small\rm 
Tensionless null p-branes in arbitrary cosmological backgrounds
are considered and their motion equations are solved. It is shown that an
ideal fluid of null p-branes may be considered as a source of gravity for
D-dimensional Friedmann-Robertson-Walker universes.
} \end{quotation}

\renewcommand{\thefootnote}{\arabic{footnote}}
\setcounter{footnote}0
\vspace{1cm}

      The string approach [1] to building a selfconsistent inflation
theory has evoked a great deal of interest now. Therefore it is important
to search the string dynamics in curved space-time [1-6]. Clearing up a
physical role of p-branes in cosmology is a topical problem, too
[4,5].  Here we consider dynamics of null p-branes [7,8], which are
tensionless limit of p-branes, in curved spaces. We show that their motion
equations can be linearized and exactly solved in contrast to the case of
p-branes. Generalizing the results of [5] we find that the perfect
fluid of the null p-branes is an alternative dominant source of gravity in
the Hilbert-Einstein equations for the Friedmann universe with $~k=0~$.
The action for null p--branes in a cosmological background
$~G_{MN}(x)~$ may be written as [7] \begin{equation}\label{1} S=\int
d^{p+1}\xi \; {det(\partial_{\mu} x^M \; G_{MN}(x) \; \partial_{\nu}
x^N)\over E(\tau,\sigma^n )}, \end{equation} where $~\mu,\nu =0,1,...,p~$
are the indices of the hyperworldsheet of the null p-brane ($~\xi^0=\tau,
\xi^1=\sigma^1, \xi^2=\sigma^2,...., \xi^p=\sigma^p~$) and
$~E(\tau,\sigma^n)~$ is a $~(p+1)~$--dimensional hyperworldsheet density.
The determinant $~g~$ of the induced null p--brane metric $~g_{\mu\nu}~$

$$    g_{\mu\nu}= \partial_{\mu}x^M\; G_{MN}(x)\; \partial_{\nu} x^N =
    \left(\matrix{\dot{x}^A \; G_{AB}(x) \;\dot{x}^B
    & \dot{x}^A \; G_{AB}(x)\;
    \partial_{n}x^B
          \cr \partial_{m}x^A \; G_{AB}(x) \;\dot{x}^B &\hat{g}_{mn}(x)\cr
	  }\right) , $$ \\
\begin{equation}\label{2}    	\hat{g}_{mn}(x) = \partial_{m}x^A \;
G_{AB}(x) \; \partial_{n}x^B \end{equation}
may be presented in a factorized form

$$ g=\dot{x}^M \;\tilde{\Pi}_{MN}(x) \;\dot{x}^N\;\hat{g} , $$
\\
\begin{equation}\label{3}
\hat{g}=\det\hat{g}_{mn} , \end{equation}
where point denotes differentiation with respect to $~\tau~$. The matrix
$~\tilde{\Pi}_{MN}(x)~$ has the properties of the projection operator [8]
\begin{equation}\label{4}  \tilde{\Pi}_{MN}=
G_{MN}- G_{MB} \;\partial_{m}x^B \;\hat{g}^{-1mn} \;\partial_{n}x^L
\;G_{LN} \end{equation} Therefore the action(1)can be written in the
following form:  \begin{equation}\label{5} S= \int d^{p+1}\xi \;
{\dot{x}^M \;\tilde{\Pi}_{MN}(x) \;\dot{x}^N \;\hat{g} \over
E(\tau,\sigma^n )}, \end{equation} The variation of the action  $~S~$ with
respect to $~E~$ generates the degeneracy condition for the induced metric
$~g_{\mu\nu}$ \begin{equation}\label{6} g=\det g_{\mu\nu}=0 ,
\end{equation}
which separates the class of $~(p+1)~$--dimensional isotropic geodesic
hypersurfaces characterized by the null volume. In the gauge
\begin{equation}\label{7}
\dot{x}^M G_{MN} \partial _m x^N=0 ; ~~~
\left({\hat{g} \over E} \right)^\bullet =0
\end{equation}
we find the motion equations
and constraints in the following form $$ \ddot{x}^M + \Gamma_{PQ}^M \;
\dot{x}^P \dot{x}^Q = 0 $$
\begin{equation}\label{8}
\dot{x}^M
G_{MN}\dot{x}^N =0,~~~~~~~~ \dot{x}^M G_{MN}\partial_{m}x^N=0
\end{equation}

Now consider the case of the $~D~$--dimensional Friedmann universe with
$~k=0~$ described by the metric form
\begin{equation}\label{9}
ds^2=G_{MN}dx^Mdx^N=(dx^0)^2-R^2(x^0) \;dx^i\delta_{ik}dx^k ,
\end{equation} where $~M, N= 0,1,...,D-1~$.
It is convenient to transform Eqs.(8)to the conformal time
$~\tilde{x}^0(\tau ,\sigma )~$, defined by \begin{equation}\label{10}
dx^0=C(\tilde{x}^0)d\tilde{x}^0,~~~~~C(\tilde{x}^0) = R(x^0),~~~~~
\tilde{x}^i = x^i
\end{equation}
In the gauge of the conformal time the
metric (9) is presented in the conformal-flat form
\begin{equation}\label{11}
ds^2=C(\tilde{x}^0) \eta_{MN}d\tilde{x}^M
d\tilde{x}^N,~~~~~~~~~ \eta_{MN}= diag(1, -\delta_{ij})
\end{equation}
with the Christoffel symbols $~\tilde{\Gamma}_{PQ}^M(\tilde{x})~~$ [9]
\begin{equation}\label{12}
\tilde{\Gamma}_{PQ}^M(\tilde{x})= C^{-1}(\tilde{x})[\delta_P^M
\tilde{\partial}_QC+\delta_Q^M \tilde{\partial}_PC -
\eta_{PQ}\tilde{\partial}^MC]
\end{equation}
Taking into account the
 relations (10,12) we transform Eqs. (8) to the form
 \begin{equation}\label{13}
 \ddot{\tilde{x}}^M +
 2C^{-1}\dot{C}\dot{\tilde{x}}^M=0
 \end{equation}
 \begin{equation}\label{14}
\eta_{MN}\dot{\tilde{x}}^M\dot{\tilde{x}}^N=0,~~~~~~~~\eta_{MN}
 \dot{\tilde{x}}^M\partial_{m}{\tilde{x}}^N=0,
 \end{equation}
where $~m,n=1,...,p~$. The first integration of these  equations leads to
the following equations of the first order
 \begin{equation}\label{15}
 H^\ast C^2\dot{\tilde{x}}^0 = \psi^0(\sigma^1,\sigma^2,..,\sigma^p),
 ~~~~~~~ H^\ast C^2 \dot{\tilde{x}}^i= \psi^i(\sigma^1,\sigma^2,..,\sigma^p ),
 \end{equation}
the solutions of which have the form
\begin{equation}\label{16}
\tau = H^\ast \psi_0^{-1} \int_{t_o}^t dt \; R(t),
\end{equation}
 $$ x^i(\tau ,\sigma^1,\sigma^2,..,\sigma^p)) = H^{\ast
 -1} \psi^i \int_0^\tau d\tau \; R^{-2}(t),$$
where $~H^{\ast}~$ is a metric constant with the dimension $~L^{-1}~$ and
$~t_0 \equiv x^0(0,\sigma^1,\sigma^2,..,\sigma^p),~$
and
$~x^i(0,\sigma^1,\sigma^2,..,\sigma^p) )~$ and $~ \psi ^M(\sigma)~$
are the initial data. The solution (16) for the space world
 coordinates $~x^i(t) (i=1,..., D-1)~$ as a function of the cosmic time $
 ~t=x^0~$, may be written in the equivalent form as
 \begin{equation}\label{17}
 x^i(t,\sigma^1,\sigma^2,..,\sigma^p)=x^i(t_0,\sigma^1,\sigma^2,..,\sigma^p)
+\nu^i(\sigma^1,\sigma^2,..,\sigma^p )\int_{t_o}^t~ dt \; R^{-1}(t),
\end{equation}
where $~\nu^i(\sigma ) \equiv \psi_0^{-1}\psi^i$.
 The explicit form of the solutions (16) allows to transform the
 constraints (11) into those for the Cauchy initial data:
\begin{equation}\label{18}
\nu^i(\sigma^m)\nu^k(\sigma^n)\delta_{ik} =1,~~
\end{equation}
\begin{equation}\label{19}
\partial_{m}x^0(0,\sigma^m)=
R(x^0(0,\sigma^m) \; \nu^i(\sigma^n)\delta_{ik}\partial_{m}x^k(0,\sigma^m),
\end{equation}
where $\nu^k(\sigma^1,\sigma^2,..,\sigma^p)=\psi^{i}
\psi_0^{-1}$. Note that the constraints (18) and (19) produce the additional
constraints , which are their integrability conditions
\begin{equation}\label{20}
\partial_{m}(\nu_i(\sigma^n)\partial_{n}x^{i}(0,\sigma^l)) -
\partial_{n}(\nu_{i}(\sigma^n)\partial_{m}{x}^{i}(0,\sigma^l))=0
\end{equation}
Now we want to show that the null p-branes may be considered as dominant
gravity sources of the Friedmann universes. With this aim assume that
the perfect fluid of these null p-branes is homogenious and isotropic.
The energy density $~\rho (t)~$ and the pressure $~p(t)~$ of this
fluid and its energy -momentum$~\langle T_{MN}\rangle~$ are connected by
the standard relations
\begin{equation}\label{21} \langle{T_ 0
}^0\rangle =\rho(t),~~~~~~~ \langle {T_i }^j \rangle = -p(t){\delta_i} ^j
= - { {\delta_i}^j \over D-1} ~{A\over R^D(t)},
\end{equation} The tensor
$~\langle T_{MN}\rangle~$ is derived from the momentum--energy tensor
$~T_{MN}~$ of a null p--brane by means of its space averaging when a set
of null p--branes is introduced instead of a separate null p--brane.  The
energy -momentum tensor $~ T^{MN}(x)~$ of null p--brane is defined by the
variation of the action (1) with respect to $~G_{MN}(x)~$
\begin{equation}\label{22} T^{MN}(X)= {1 \over \pi\gamma^\ast \sqrt{
_{|G|}}} \int ~d\tau d^{p}\xi \; \dot{x}^M\dot{x}^N\delta^D(X^M-x^M)
\end{equation} After the substitution of the velocities $\tilde{x}^m$
(15) and subsequent integration with respect to $~\tau~$, the non-zero
components of $~T_{MN}~$ (22) take the following form $$
T^{00}(X)=\frac{1}{\pi\gamma^* H^*}R^{(-D)}(t)\int d\tau d^{p}{\xi} \
 \psi_0(\sigma^{m})\delta^{D-1}(X^i-x^i(\tau,\sigma^{m})), $$
\begin{equation}\label{23}
T^{ik}(X)=\frac{1}{\pi\gamma^*H^*}R^{(-D-2)}\int d\tau {d^{p}\xi} \
{\nu^i(\sigma)\nu^k(\sigma)\psi_0(\sigma)\delta^{D-1}(X^i-x^i(\tau,\sigma^{m})}
,\end{equation}
where the time dependence $~T^{MN}~$ is factorized and
accumulated in the scale factor $~R(t).~$  Taking into account the
constraint$~\nu^i(\sigma^m)\nu^k(\sigma^n)\delta_{ik} =1~~$ gives rise
to the following relation between the components of the tensor $~\langle
T_{MN}\rangle~$
\begin{equation}\label{24}
Sp~T ={T_0}^0+G_{ij}T^{ij}=0 .
\end{equation}
As a result of the space averaging we find the non--zero components of
$~\langle T_{MN}\rangle~$ to be equal to
\begin{equation}\label{25}
\langle {T_0 }^0 \rangle = \rho(t) =  {A \over R^D(t)},~~~~~~
\langle {T_i }^j \rangle = -p(t){\delta_i} ^j ,
\end{equation}
where $~A~$ is a constant with the dimension $~L^{-D}~$.  Eqs.(25) show that
the equation of state of null p--branes fluid is just the equation of
state for a gas of massless particles \begin{equation}\label{26} \langle
Sp~T\rangle = \langle {T_M} ^M \rangle =0 \Longleftrightarrow \rho =
(D-1)p
\end{equation}
Eq. (26) was found in [3]as the approximate equation of state describing
the phase of perfect gas of shrunk strings $~(R(t)\rightarrow 0,~ \tau
\to0)~$ valid in the small $~R(t)~$ limit for a negatively accelerated
contraction $(d^2R/dt^2 < 0, ~ dR/dt < 0)~$ of the universes. Taking into
account the additional interactions of the null p--brane fluid with any
background fields breaks the coincidence of their state equation and
the one describing ultrarelativistic gas of massless particles in the same
background. It is a consequence of the fact that particles have no internal
structure in contrast to extended objects such as p--branes. An example
of such an interaction may be the interaction of null p--brane with the
dilaton field $~\phi~$  describing by the action [7]
\begin{equation}\label{27} S=S_0+S_1 = \int d\tau d^{p}\xi\left[{det
(\partial_\mu x^M G_{MN} \partial_\nu x^N)\over E(\tau,\sigma^m)} -
{\phi(x^M)} E (\tau,\sigma^m)\right]~~, \end{equation}

 Now assume that the fluid of null p--branes is a dominant source of the FRW
 gravity (1).  For the validity of the last conjecture it is
 necessary that the H--E equations
 \begin {equation}\label{28}
 {R_M} ^N =
8\pi G_D\langle {T_M}^N \rangle
\end{equation}
with the non-zero Ricci
tensor $~{R_M} ^N~$ components defined by $~{G_M} ^N~$ (1) $$~ {R_0} ^0 =
-{D-1 \over R}~{d^2R\over dt^2}, $$
\begin{equation}\label{29}
{R_i} ^k =
-{\delta _i} ^k \left[{1\over R}~ {d^2R\over dt^2} + {D-2\over R^2} \left(
{dR\over dt}\right)^2\right]
\end{equation}
should contain the tensor
$\langle{T_M} ^N\rangle$ (21) as a source of the FRW gravity.  Moreover,
the constraints (21), i.e. $$ \rho R^D - A = 0 $$ must be a motion
integral for the HE system (28). It is actually realized because $$
{d\over dt}(\rho R^D) = - {D-2\over 16 \pi G_D } R^{D-1}~{dR\over dt }
{R_M} ^M = 0, $$ since the trace  $~{R_M}^M \sim \langle{T_M}^M\rangle=0~
$ (see (26). In view of this fact it is enough to consider only one
equation of the system (28)
\begin{equation}\label{30}
\left({1\over
R}~{dR \over dt}\right)^2 = {16 \pi G_D \over(D-1)(D-2)} ~{A\over R^D}\ \
\end{equation}
which defines the scale factor R(t) of the FRW metric
(1). Note that in the case  $D~=~4$ Eq.(30) transforms into the well-known
Friedmann equation for the energy density in the radiation dominated
universe with  $~k = 0~$.  The solutions of Eq. (30) are $$
R_I(t) = [q(t_c -t)]^{2/D}, ~~~~~~~ t < t_c, $$ $$ R_{II}(t) =[q(t
-t_c)]^{2/D},~~~~~~ t > t_c , $$ where $q=[4\pi G_D A /
(D-1)(D-2)]^{1/2}~~$ and $t_c$ is a constant of integration which is a
singular point of the metric. The solution $~R_I~$ describes the stage
of negatively accelerated contraction of D-dimensional FRW universe. In
the small $R$ limit $(R\rightarrow 0)$ [3] this solution was found as an
approximate asymptotic solution for the gas of strings with non-zero
tension. For the case of null strings [5]  and null p-branes this
solution is exact.  The second solution (30) $~R_{II}~$ describes the
stage of the negatively accelerated expansion of the FRW universe from the
state with space volume equal to "zero".  Thus we see that the perfect
fluid of noninteracting null p-branes may be considered as an
alternative source of the gravity in the FRW universes with $~k = 0~$.

Null p--branes theory in curved space--time is characterized by a set of
the constraints connected with the reparametrization symmetry. To find
these constraints consider the canonical momentum of null p--brane
$~{\cal P}_M~$ conjugated to its world coordinate $~x_M~$
\begin{equation}\label{31}
{\cal P}_M = 2E^{-1}\; \hat{g} \;\tilde{\Pi}_{MN}(x) \;\dot{x}^N
\end{equation}
Then we find the following primary constraints
\begin{equation}\label{32}
 G_{MN} \partial _m x^N {\cal P}^M = 0  ~~~
\end{equation}
The Hamiltonian density produced by the action functional (1) is
\begin{equation}\label{33}
{\cal H}_0 = {1\over 4}E \;\hat{g}^{-1} G^{MN} {\cal P}_M{\cal P}_N
\end{equation}
and the condition of conservation of the primary constraint
\begin{equation}\label{34}
		{\cal P}_{(E)} = 0,
\end{equation}
where $~{\cal P}_{(E)}~$ is the canonical momentum conjugated to $~E,~$
generates the following condition
\begin{equation}\label{35}
\dot{\cal P}_{(E)} = {\int d^{p}\xi  \;\{\cal H}_{0} ,{\cal
P}_{(E)}\}_{P.B.} = - {1\over{4\hat g}}G^{MN}{\cal P}_M{\cal P}_N = 0
\end{equation} produces a secondary constraint \begin{equation}\label{36}
G^{MN}{\cal P}_M{\cal P}_N = 0
\end{equation}
Additional constraints do not appear so the total hamiltonian of null
p-brane is given by
\begin{equation}\label{37}
 H = \int d^{p}\xi \  [\lambda^{m}(G_{MN} \partial _m x^N {\cal P}^M) +
{E\over{4\hat g}}G^{MN}{\cal P}_M{\cal P}_N  + \omega{\cal P}_{(E)}]   ,
\end{equation}
This hamiltonian and the reparametrization constraints may be used for the
quantization of null p--brane in a curved space time.

This paper was supported in part by the INTAS and Dutch Government Grant
 94--2317 and the Fund for Fundamental Research of CST of Ukraine No 2.3/664.

\newpage 

\vspace{15mm}
 \begin{center} REFERENCES
 \end{center}

\begin{enumerate}
\item G.Veneziano, Proceedings of the Erice Course "String Gravity and
Physics at the Planck Scale"edited by N.Sanchez and A.Zichichi NATO ASI
Series,C476,1996.  \\ M.Gasperini,  Ibid.  \item H.J.De Vega and
N.Sanchez, Ibid.  \\ A.L.Larsen and N.Sanchez , Ibid.
\item N.Sanchez and G,Veneziano,
Nucl.Phys.{\bf B333} (1990) 253.  \\ M.Gasperini, N.Sanchez and
G.Veneziano, Nucl.Phys. {\bf B364} (1991) 265.  \item M.J.Duff, R.P.Khuri
and J.X.Lu , Phys.Repts. {\bf 259} (1995) 213.   \item S.N.Roshchupkin and
A.A.Zheltukhin, Class.Quant.Grav. {\bf 12} (1995) 2519. \item S.Kar,
"Schild null string in flat and curved background", Preprint IP,
Bhubanesvar,1995   \item A.A.Zheltukhin, JETP Lett.{\bf 46} (1987) 208. \\
Sov.J.Nucl.Phys. {\bf 48}(1988) 375; {\bf 51} (1990) 1504.  \item
I.A.Bandos and A.A.Zheltukhin, Fortschr.Phys.  {\bf 41} (1993) 619.  \item
A.Z.Petrov, " New Methods in General Relativity", Nauka,Moscow, 1996.
\end{enumerate}

\end{document}